\documentclass[aps,prl,superscriptaddress,twocolumn,amsmath,amssymb,showpacs,floatfix,10pt]{revtex4-1}

\usepackage{amsmath}
\usepackage{amssymb}
\usepackage{graphicx}
\usepackage{xspace}		

\usepackage{dcolumn}		
\usepackage{bm}			
\usepackage{subfigure}	
\usepackage{hyperref}	
  \hypersetup{colorlinks = true, 	
  citecolor = blue,		
  linkcolor = red, 		
  breaklinks = true}

\usepackage{upgreek}

\newcommand{\eg}{{\sl e.g.\xspace}}
\newcommand{\ie}{{\sl i.e.\xspace}}
\newcommand{\vo}{V$_\textrm{O}$\xspace}

\newcommand{\ozc}{O$_\textrm{2c}$\xspace}

\newcommand{\tifc}{Ti$_\textrm{5c}$\xspace}
\newcommand{\tiint}{Ti$_\textrm{int}$\xspace}
\newcommand{\ana}{anatase (101)\xspace}
\newcommand{\tio}{TiO$_{2}$\xspace}

\makeatletter

\begin{document}
\title{(Sub)surface mobility of oxygen vacancies at the \tio \ana surface}

\author{Philipp Scheiber}
\affiliation{Institute of Applied Physics, Vienna University of Technology, Wiedner Hauptstrasse 8-10/134, 1040 Vienna, Austria}

\author{Martin Fidler}
\affiliation{Institute of Applied Physics, Vienna University of Technology, Wiedner Hauptstrasse 8-10/134, 1040 Vienna, Austria}

\author{Olga Dulub}
\affiliation{Department of Physics, Tulane University, New Orleans, LA 70118, USA}

\author{Michael Schmid}
\affiliation{Institute of Applied Physics, Vienna University of Technology, Wiedner Hauptstrasse 8-10/134, 1040 Vienna, Austria}

\author{Ulrike Diebold}
\email{diebold@iap.tuwien.ac.at}
\affiliation{Institute of Applied Physics, Vienna University of Technology, Wiedner Hauptstrasse 8-10/134, 1040 Vienna, Austria}
\affiliation{Department of Physics, Tulane University, New Orleans, LA 70118, USA}

\author{Weiyi Hou}
\affiliation{Department of Chemistry, Princeton University, Frick Laboratory, Princeton NJ 08544, USA}

\author{Ulrich Aschauer}
\affiliation{Department of Chemistry, Princeton University, Frick Laboratory, Princeton NJ 08544, USA}

\author{Annabella Selloni}
\affiliation{Department of Chemistry, Princeton University, Frick Laboratory, Princeton NJ 08544, USA}

\date{\today}


\begin{abstract}

Anatase is a metastable polymorph of  \tio. In contrast to the more widely-studied  \tio rutile, O vacancies (\vo's) are not stable at the \ana surface. Low-temperature STM shows that  surface \vo's, created by electron bombardment at 105\,K, start migrating to subsurface sites at temperatures  $\geq$ 200\,K. After an initial decrease of the \vo density, a temperature-dependent dynamic equilibrium is established where \vo's move to subsurface sites and back again, as seen in time-lapse STM images. We estimate that activation energies for subsurface migration  lie between 0.6 and 1.2\,eV; in comparison, DFT calculations predict a barrier of ca. 0.75 eV. The wide scatter of the experimental values might be attributed to inhomogeneously-distributed subsurface defects in the reduced sample.  
\end{abstract}



\maketitle



Titanium dioxide, \tio, is one of the most versatile oxide materials and finds wide use, \eg, in energy-related applications such as (photo-)catalysis and solar energy conversion schemes.   \tio has also evolved as a popular model system for studying the fundamentals of defect-related surface processes at the molecular scale \cite{diebold_surface_2003, pang_chemical_2008}.

 \tio crystallizes in three different structures commonly named rutile
($D^{14}_{4h}$--$P4_{2}/mnm$), anatase 
($D^{19}_{4h}$--$I4_{1}/amd$), and brookite 
($D^{15}_{2h}$--$Pbca$).
\tio nanomaterials can be synthesized with various shapes and functionalities using sol-gel and other processing techniques \cite{Chen_Titanium_2007}. Although the anatase polymorph is metastable, it is commonly found in nanomaterials where the crystal size is below a few tens of nm. Yet  few experimental studies on large single crystals exist \cite{Diebold_One_2003, hebenstreit_scanning_2000, Ruzycki_Scanning_2003, Xu_photocatalytic_2011, Walle_mixed_2011}, thus the surfaces of anatase are not as well understood as those of rutile, where processes related to intrinsic defects -- Ti interstitials (\tiint) and surface O vacancies (\vo) -- have received considerable attention \cite{dohnalek_thermally_2010, Wendt_role_2008, Yim_Oxygen_2010, zhang_direct_2010}.

Recently we have found a significant difference between the surfaces of rutile and anatase: at anatase (101), the most stable surface of this polymorph, it is energetically more favorable for O vacancies to reside in the bulk than  on the surface \cite{he_evidence_2009}.  This is in stark contrast to rutile (110), where surface \vo's form easily under standard preparation conditions  \cite{diebold_surface_2003}. The preponderance of bulk defects in anatase was first predicted by DFT calculations, which showed that the formation energy of a  surface \vo is larger than that of a bulk vacancy by about $\sim$0.5\,eV  \cite{cheng_energetics_2009, cheng_surface_2009}. In a previous STM study \cite{he_evidence_2009} we compared a freshly-cleaved, pristine anatase (101) sample with a more O-deficient, reduced one.  STM images of the reduced \ana surface have an inhomogeneous appearance  that strongly depends on the STM tunneling parameters; we attributed this to a variation of the local electronic structure due to subsurface defects, i.e., O vacancies and/or Ti interstitials.  We also found that more reduced anatase is more reactive towards water adsorption, despite the fact that no \vo's are visible at the surface \cite{aschauer_influence_2010}. 

The observation that surface \vo's are less stable than bulk \vo's is remarkable. An O atom can leave a solid only through its surface, thus an as-formed surface \vo should diffuse into the bulk. The activation energy (E$_{act}$) for surface-to-subsurface migration is $\sim$ 0.7\,eV according to our DFT calculations.  Such surface-to-bulk migration should thus be observable at temperatures that are conveniently accessible in an STM experiment; this work presents such a study. We create surface \vo's non-thermally by electron bombardment \cite{dulub_electron-induced_2007}, and monitor their fate with low-temperature and variable-temperature STM. We find that surface \vo's diffuse to subsurface sites at temperatures above 200\,K. Time-lapse STM images show a temperature-dependent, dynamic equilibrium concentration of surface defects. The results point towards an activation energy for subsurface migration of a \vo that depends on its immediate surroundings.

\begin{figure}[h!tb]
  \centering
    \scalebox{0.49}{\includegraphics{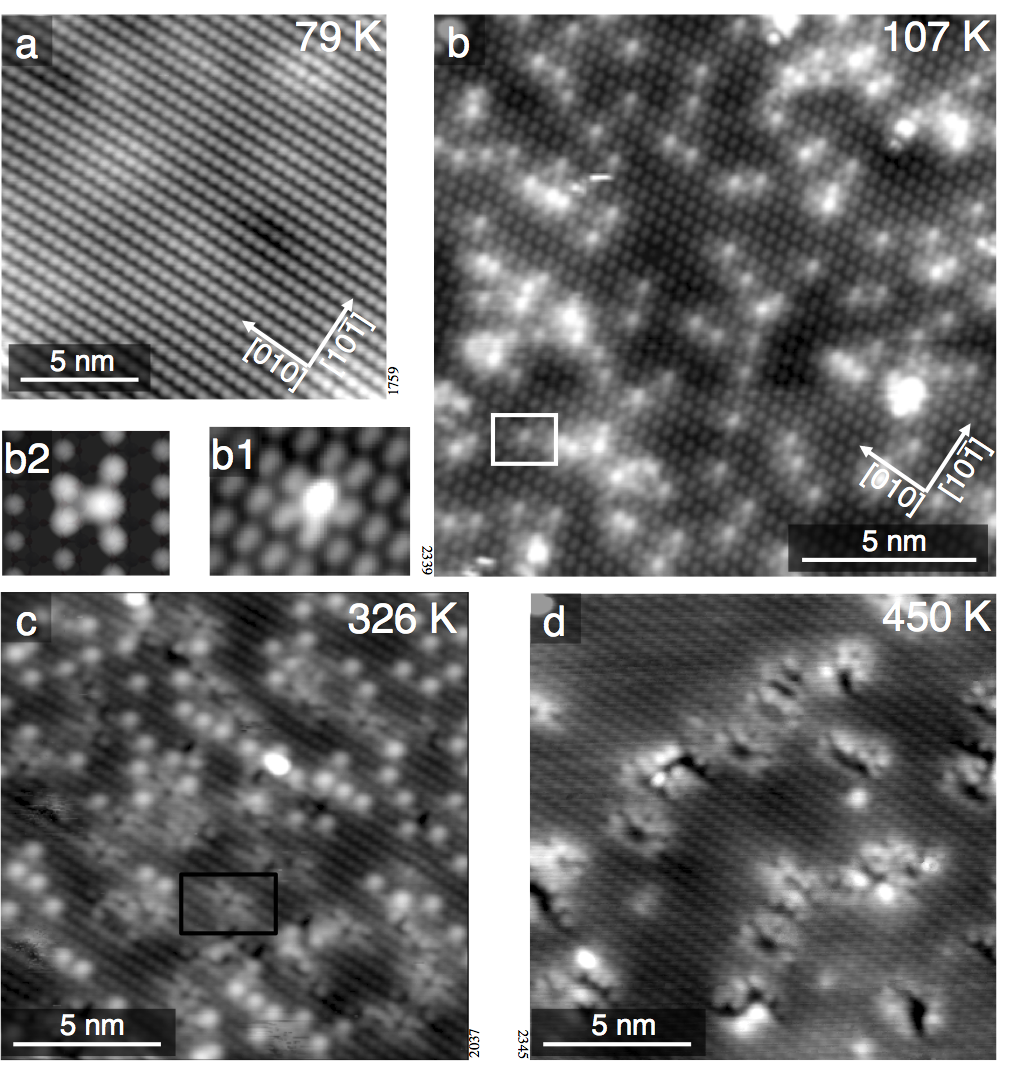}}
    \caption{STM images (T$_{sample}$ = 78\,K) of \tio \ana. (a) Freshly-prepared surface. (b) After              irradtion with 500\,eV electrons, which creates surface O vacancies (\vo's). The insets (b1, b2) show a magnified experimental and calculated STM image of a \vo, respectively. After annealing the sample for 10 minutes to (c) 326\,K     and (d) 450\,K}
  \label{Fig1} 
  \end{figure}

\begin{figure}[htb!]
  \centering
    \scalebox{0.8}{\includegraphics{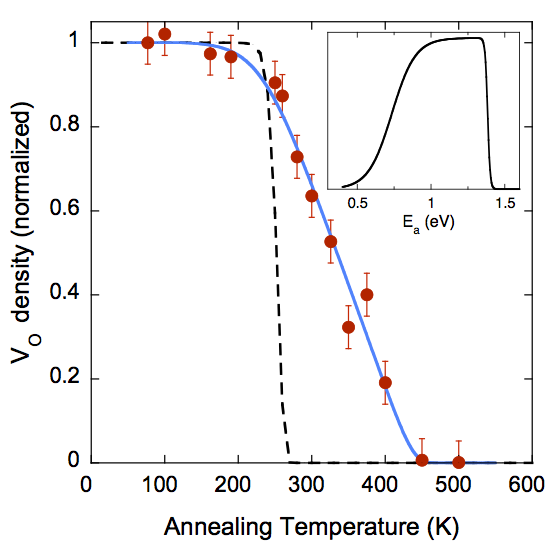}}
    \caption{(Color online) Stability of surface \vo's created by electron beam bombardment. The plot shows the density of         \vo's after heating the sample to various temperatures for 10 minutes, normalized to the initial value after electron  irradiation at 105\,K. The dashed line shows the expected behavior assuming one E$_{act}$ of 0.75\,eV. The full line        assumes the trapezoidal distribution of E$_{act}$'s from 0.6 -- 1.2\,eV displayed in the inset.}
 \label{Fig2}   
\end{figure}

The experiments were carried out in a two-chamber UHV system with a base pressure of 10$^{-11}$\,mbar. Unless noted otherwise, constant current STM measurements were performed  at 78\,K.  For STM we typically used positive sample bias voltages between 1.3 and 1.5\,V, and tunneling currents between 0.1 and 0.4\,nA for STM.   
A mineral \ana sample was cleaved \textit{ex-situ} as described in reference \cite{dulub_preparation_2010}. A clean, almost pristine surface was repeatedly prepared by sputtering (1\,keV Ar$^+$, fluence of 6.7 $\times$ 10$^{15}$ ions/cm$^2$), annealing in O$_2$ (p = 5 $\times$ 10$^{-7}$\,mbar) at 923\,K for 30 minutes, and post-annealing in UHV at 973\,K for another 10 minutes, see Fig.\ref{Fig1}(a). To create \vo's, the surface was irradiated with a rastered and thoroughly-outgassed electron gun at a current density of 1\,$\mu$A\,mm$^{-2}$ (current measured with a positive sample bias of 27\,V). Electron bombardment  was performed in the preparation chamber with the sample kept at 105\,K.  As is shown below, \vo's are immobile at this temperature.   After irradiation the sample was transferred into the STM for analysis.  To determine the stability of the electron-induced surface defects (Fig. \ref{Fig2}),  we proceeded as follows:  The manipulator in the preparation chamber was resistively heated and equilibrated at the desired temperature.  With a pre-cooled wobblestick the sample was taken from the cold STM and inserted into the manipulator, where it was kept for 10 minutes.  Then the sample was transferred back into the cold STM. The minimum time between taking the sample from the manipulator and the first usable STM image was also 10 minutes.  It is important to note (see below) that the initial \vo density was kept constant throughout these experiments.

The DFT calculations were performed using the Perdew-Burke-Ernzerhof (PBE) \cite{Perdrew_generalized_1997} functional and the plane wave pseudopotential scheme as implemented in the  Quantum ESPRESSO package \cite{giannozzi_quantum_2009}. In addition, selected spin polarized hybrid PBE0 calculations \cite{Perdew_1996} were performed using a mixed localized + plane wave  basis set expansion of the electronic states  as implemented in CP2KQuickstep    \cite{vandevondele_quickstep_2005}.
The defected surface was modeled using 3$\times$1 (10.26$\times$11.31  \AA$^2$) supercells with periodically repeated slabs of three (9.7\,\AA) or four (13.1\,\AA) \tio layers separated by a vacuum of about 10 \AA. For STM calculations, larger 4$\times$2 (20.49$\times$15.06\,\AA$^2$) supercells were used to separate the periodic images. Activation energy barriers were estimated using the Nudged Elastic Band (NEB) \cite{henkelman_climbing_2000, Mills_Quantum_1994} method. Other computational details are given in the Supplemental Material.


The sputtered and annealed \ana surface is characterized by trapezoidal islands; their orientation indicates the crystallographic directions of the crystal \cite{Gong_steps_2006}. Atomically-resolved STM shows rows of oval-shaped spots that extend over both, the \tifc and \ozc surface atoms \cite{hebenstreit_scanning_2000}, oriented along the [010] direction, (see Fig. \ref{Fig1}(a)). Our sample preparation procedure renders a bulk-reduced sample, as evidenced by a small shoulder in the XPS Ti2p core levels. The surface has a non-uniform appearance in STM, with a long-range corrugation that depends strongly on the tunneling conditions as observed previously  \cite{he_evidence_2009}; these are attributed to either intrinsic or extrinsic subsurface defects.  

\tio is sensitive to electron irradiation, which can be used to create vacancies at the undercoordinated O sites of the surface  \cite{dulub_electron-induced_2007, Pang_Tailored_2006}. An STM image of an electron-irradiated \ana surface is shown in Figure \ref{Fig1}(b). \vo's appear as extra bright features at regular lattice sites, consistent with STM simulations, see the Figs. \ref{Fig1} (b1, b2).  After exposure to 6.6 $\times$ 10$^{14}$ electrons/cm$^2$,  the density of  such \vo's amounts to 12\,\% of a monolayer (ML, where 1 ML is defined as the number of primitive unit cells, \ie, 3.8 $\times$ 10$^{13}$\,cm$^{-2}$). Assuming a simple, first-order desorption process, we estimate a cross section for electron-induced O desorption  of 3 $\times$ 10$^{-19}$\,cm$^{-2}$.

The stability of these surface vacancies was probed by annealing the electron-irradiated sample for 10 minutes as described above.  Each heating excursion was performed with a freshly-prepared and irradiated surface; the \vo densities after the annealing steps are shown in Fig. \ref{Fig2}. No significant change was observed up to a temperature of 200\,K;  after an anneal to   ~230\,K, the defect density decreases significantly.  The higher the sample temperature during the 10 min anneal, the fewer \vo's survive. Above 320 K, new features appear that span several unit cells, one is marked with a black box in Fig. \ref{Fig1}(c).  These features (not taken into account in Fig.\ref{Fig2})  become more extended when an electron-irradiated surface is heated to higher temperatures (Fig. \ref{Fig1}(d)), and disappear completely above 500 K. 

\begin{figure}[htb!]
  \centering
    \scalebox{0.8}{\includegraphics{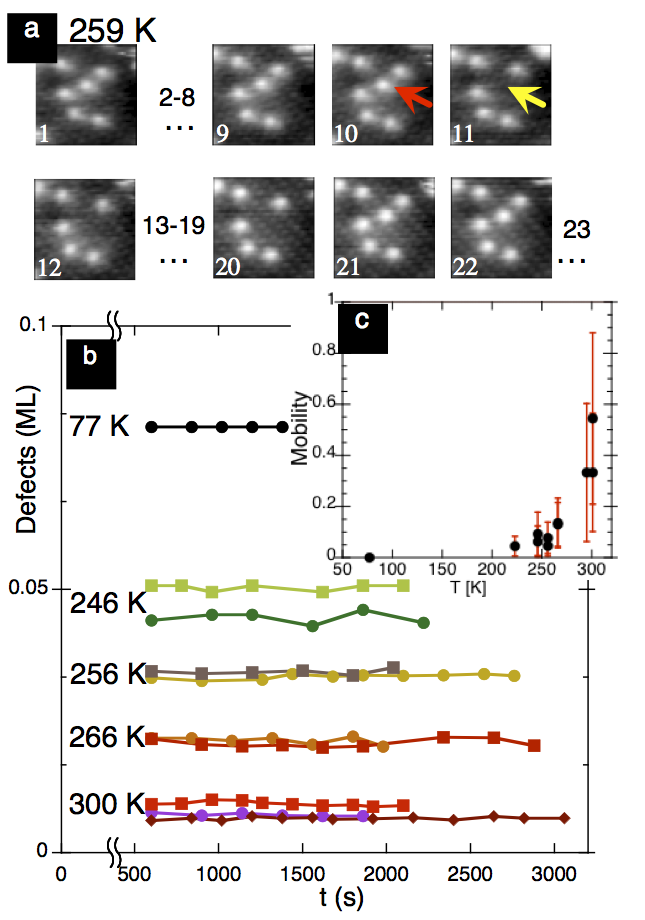}}
    \caption{(Color online) Results from time-lapse STM images of surface \vo's on \ana. (a) Series of images (4 $\times$          4\,nm$^2$; +1.6\,V / 0.2\,nA) recorded at T = 259\,K; the time between images was 3.2 minutes. The arrows mark a \vo that          disappears and re-appears at the same position. (b) Total defect density in time-lapse images; each trace corresponds to a separate experimental run at the sample temperature indicated. (c) Defect mobility,  represented by the number of hopping events per defect and frame for different temperatures.} 
  \label{Fig3}   
\end{figure}

In addition to heating excursions, we also followed the fate of single \vo's in time-lapse images at various temperatures.  For these measurements we first equilibrated the STM for several hours at a specific temperature between 220 and  300\,K.  Electron bombardment of the freshly-prepared sample was again performed at 105 K. (At this temperature we do not expect any surface-to-bulk migration, Fig.  \ref{Fig2}.) The irradiated sample was inserted into the temperature-stabilized STM, and series of images were taken.  Fig. \ref{Fig3}(a) shows an example of such a time-lapse sequence, taken at T$_{sample}$ = 259\,K.  One of the defects, marked with an arrow, disappears and returns to the same spot a few frames later.  We also observed that defects disappeared at one position and appeared at another position at the same or -- less frequently -- a neighboring row.  The mobility of \vo's increases with temperature, see Fig. \ref{Fig3}(c).  The total defect density, however, remains constant within the time frame of the experiment, see Fig. \ref{Fig3}(b). 

It takes at least 10 minutes between the end of electron-irradiation (at 105\,K) and the recording of the time-lapse sequences in our experimental setup.  During this time the total defect density decreases significantly, as shown in Fig. \ref{Fig2}.  This is the reason why the absolute \vo densities in Fig. \ref{Fig3}  vary with temperature. On the other hand, the fact that the number of defects stays constant (Fig. \ref{Fig3}(c)) after the original, rapid decrease gives us confidence that the data displayed in Fig. \ref{Fig2} indeed show the equilibrium concentrations at the given temperatures, and that the finite time constants of our experiment do not influence the results.


By DFT calculations, we estimate that the barrier, E$_{act}$, for surface-to-bulk migration of \vo's is 0.75\,eV, while it is  1.15 \,eV for the reverse process (slight differences with respect to the barriers in Ref. \cite{cheng_surface_2009} are due to the larger surface model used for the present NEB calculations). The dashed line in Fig. 2 shows the expected behavior if we adapt this E$_{act}$ and a conventionally-used pre-factor of 10$^{12}$\,s$^{-1}$.  While the onset of bulk migration is consistent with the DFT result, the expected decrease with temperature is much steeper than the measured one. In addition, the  dis/re-appearance of the surface \vo's, which leads to a temperature-dependent, dynamic equilibrium is hard to reconcile with the picture derived from our DFT calculations: once the sample temperature is high enough to overcome the energetic barrier for surface-to-bulk migration, there is little reason for a \vo to return back to the surface. One should consider, however, that the calculations were performed assuming an idealized case, \ie, a perfect anatase slab devoid of any other defects except the single \vo under investigation. This is different from the situation in the experiment, where subsurface defects are present at the outset. From titration experiments using O$_2$ adsorption we estimate that the density of \tiint's and \vo's in the near-surface region of our sample amounts to 2 ($\pm$ 1)\,\% of a ML at the clean, as-prepared surface.  The uneven appearance of the STM images from the clean surface (Fig. \ref{Fig1}(a)) is attributed to local band bending effects.  Thus at least some of the subsurface defects are charged; plausibly these exert a considerable influence on the energetics and dynamics of defects migrating within their neighborhood. It is not unreasonable to assume a range of E$_{act}$'s for subsurface diffusion, as this value will depend on the immediate environment of each surface \vo. The full line in Fig. \ref{Fig2} takes into account such a scenario, where we assume a trapezoidal distribution of E$_{act}$'s ranging from 0.6 to 1.2\,eV, as displayed in the  inset of Fig. \ref{Fig2}).  

The time- and temperature dependent behavior of \vo's can also be explained with such a range of activation energies: starting with a certain surface \vo concentration, the defects that happen to reside above relatively perfect region of the sample can disappear into the bulk once a temperature  $>$ 200\,K is reached.  If another defect is present within the selvedge of the crystal, it will affect the \vo and change the activation energy for its disappearance into the bulk. It is well possible that the defect migrates a certain distance in the subsurface region before it pops up again -- estimates for lateral diffusion energies are in the range  1.1 - 1.8 \,eV (see Supplemental Material), hence the \vo's can appear at different positions, as is observed in the experiment. The extended features observed in Figs. \ref{Fig1}(c, d) suggest that \vo's aggregate in the near-surface region at moderate annealing temperatures. The temperature dependence of bulk diffusion and defect equilbria observed in this work are possibly affected by the initial  \vo concentration; this  could be tested in future experiments.

The experimental results presented in this work are unequivocal proof for the  theoretical prediction that vacancies are more stable in the bulk than at the surface. 
This prediction, originally based on DFT-PBE calculations  \cite{cheng_energetics_2009, cheng_surface_2009}, is also supported by results from hybrid calculations which account for the polaronic character of \vo-induced Ti$^{3+}$ states and are thus considered more accurate for the study of defects in  \tio    \cite{Ganduglia_oxygen_2007, Finazzi_excess_2008}, see Supplementary Material.  While  hybrid calculations are still too demanding to be used for diffusion barrier determinations, DFT+U  studies indicate that the barriers for the hopping diffusion of the Ti$^{3+}$ polarons are low, typically  between 0.1 and 0.3 eV
\cite{deskins_electron_2007,kowalski_charge_2010,deskins_distribution_2011}
Therefore the effect of excess electron localization on \vo migration barriers is expected to be relatively minor, as has recently been shown for H diffusion in anatase.
\cite{aschauer_subm_2012}   

An inspection of the anatase (101) surface structure provides a simple qualitative rationale for the instability of surface \vo's: removal of an \ozc gives rise to one five-fold and one highly unstable four-fold coordinated Ti$^{3+}$ cation, whereas bulk \vo's have two five-fold coordinated Ti$^{3+}$ cations.
Moreover, the Ti-\ozc bonds are short and strong, so breaking two Ti-O bonds at the surface is energetically more costly than to breaking three in the bulk.  Clearly, the resulting subsurface defects have to be reckoned with when considering the surface chemistry of \tio anatase, and some observations have already been interpreted along these lines \cite{aschauer_influence_2010, Xu_dissociation_2012}. Subsurface migration automatically results in inhomogeneity within the selvedge of the crystal, which, in turn, affects the activation energies.  The dynamic equilibrium of surface O vacancies will then depend on the presence of intrinsic as well as extrinsic charged defects.  Even at room temperature defects come and go from the surface, suggesting that the chemically active sites change across the surface. 

Generally, the flow of lattice oxygen (defects) to and from the surface is of continued interest in solid-state chemistry, and important in established and emerging technologies such as catalysis \cite{Esch_ceria_2005},  solid-oxide fuel cells \cite{Suntivich_design_2011} and memristor devices \cite{Waser_nanoionics_2007}. Direct observation of such defect migration, combined with modeling at the atomic scale can help pave the way for future experiments that give insights into the relevant processes.  

Acknowledgement: This work was supported by the Austrian Science Fund (FWF; Project F45) and the ERC Advanced Grant 'OxideSurfaces'.
AS acknowledges support from DoE-BES, Chemical Sciences, Geosciences and Biosciences Division under Contract No. DE-FG02-12ER16286. Calculations were performed  at the TIGRESS high performance computer center at Princeton University.
\bibliography {bibliography}

\end{document}